
\line{\hfil TAUP 2110-93}

\def \ra {\rightarrow}
\def \p {{\cal P}}

\medskip
\titlepage{A-B AND BERRY PHASES FOR A QUANTUM CLOUD OF CHARGE}
\vskip 2 true cm
\centerline{
{{Yakir Aharonov,\footnote{a}{School of Physics and Astronomy,
Tel-Aviv University, Ramat-Aviv 69978 Israel}}\footnote{b}{also
Department of Physics and Astronomy, University of South Carolina,
Columbia SC 29208}}
{Sidney Coleman,\footnote{c}{Lyman Laboratory, Harvard University,
Cambridge, MA 02138}}
{Alfred S. Goldhaber,\footnote{d}{Institute for Theoretical Physics,
State University of New York, Stony Brook, NY 11794-3840}}
{Shmuel Nussinov,$^a$}
}
\centerline{
{Sandu Popescu,\footnote{e}
{Service de Physique Theorique, Universit\'e Libre de Bruxelles,
Campus Plaine, C.P. 225, Boulevard du Triomphe, B-1050 Brussels,
Belgium}}
{Benni Reznik,$^a$}
{Daniel Rohrlich,$^a$}
and {Lev Vaidman$^a$}
}
\vskip 4 true cm

\abstract
We investigate the phase accumulated by a charged particle in an extended
quantum state as it encircles one or more magnetic fluxons, each carrying half
a flux unit. A simple, essentially topological analysis reveals an interplay
between the Aharonov-Bohm phase and Berry's phase.
\vfill
\eject

     The Aharonov-Bohm (AB) phase\Ref\ab{Y.~Aharonov and D.~ Bohm, {\it Phys.
Rev.} {\bf 115}, 485 (1959).} $\Phi_{AB}= (q /\hbar )\oint_C\vec A\cdot d\vec
r$ collected by a charge $q$, moving in a closed path $C$ about a line of
magnetic flux $\phi$, is purely topological: $\Phi_{AB}= 2\pi n (q/e)(\phi
/\phi_0)$, with $n$ the winding number of $C$ around the fluxon, $e$ the
elementary charge, and $\phi_0$ the corresponding flux unit $\phi_0={2\pi\hbar
/e}$. The AB phase is independent of the shape of the path $C$ and of the
history of motion along it.  If the charge is not pointlike, or the fluxon is
not linelike, they may overlap; what then happens to the AB phase? As long as
the charge and flux are distributed classically, the answer is straightforward:
 a system of charges moving in a closed path through a classical magnetic field
collects an AB phase $\Phi_{AB}= (1/\hbar ) \sum q_i\phi_{i}$, with $\phi_{i}$
the flux enclosed by the path of the $i$-th charge. Here, however, we consider
charges distributed by quantum smearing. The phase of a {\it quantum} charge is
no simple sum over the undeformed charge distribution.  If we compute it via
the Born-Oppenheimer approximation, we find a remarkable interplay between the
AB phase and Berry's phase that determines the overall topological phase.

     Consider a single electron bound to a heavy ``nucleus'' (assumed neutral,
for simplicity) in the presence of an infinitely long flux line. Both the
nucleus and the fluxon may move. If the fluxon makes a closed path around the
nucleus, it may encircle some parts of the quantum charge distribution (the
electron ``cloud'') and not others.  Suppose that the time in which the fluxon
crosses the ``atom", multiplied by typical electronic frequencies, is much
smaller than 1.  In this limit, the initial electronic wave function $\Psi_0$
transforms into $\Psi_1 = e^{i\Phi_{AB}}\Pi_C \Phi_0 + (1-\Pi_C ) \Phi_0$,
where $\Pi_C$ projects onto the part of $\Psi_0$ that the fluxon encircles.
The electron has no time to move between the two parts of $\Psi_0$. But, except
in this limit, we cannot assign parts of the wave function distinct phases.
Consider now the opposite limit, of adiabatic motion. In this limit, another
phase effect comes into play.  Berry's phase\Ref\be{M.~V.~Berry, {\it Proc.
Roy. Soc. London} {\bf A392}, 45 (1984).} arises when parameters for a quantum
system vary adiabatically in a closed path. Applying the Born-Oppenheimer
approximation to the fluxon-atom system, and for definiteness fixing the
nucleus, we obtain {\it both} an AB phase and a Berry phase. An AB phase arises
from motion of the fluxon with respect to the instantaneous charge
distribution, while a Berry phase arises from rearrangement of the electronic
wave function.  There is a subtle interplay of these two phases, which is
purely topological for special values of the flux $\phi$ carried by the fluxon.
 For example, when $\phi=n\phi_0$, the phases completely cancel \Ref\prep{B.~
Reznik and Y.~Aharonov, {\it Phys. Lett.} {\bf B315}, 386 (1993).} as expected
since the fluxon is a pure gauge artifact. Here, we consider the more
interesting case of fluxons carrying half a flux unit (``half-fluxons" or
``semifluxons"). Topological analysis, with no computations, reveals the
interplay of the AB phase and Berry's phase.

     Let us begin with the electron (mass $m_1$) bound at the origin with a
potential $V(\vec r_1 )$ and the fluxon (mass $m_2$) free to move but
constrained to remain parallel to the $z$-axis.  The Hamiltonian is
$$
H= {(\vec p_1 +e\vec A)^2\over 2m_1 }+V(\vec r_1 )
+ {(\vec p_2 -e\vec A)^2\over 2m_2 }~~~~;
\eqn\hamhalf $$
for a half-fluxon we take $\vec A = (\hbar /2e) \vec \nabla_2 \varphi_{12}$
with $\varphi_{12}$ the angle of the fluxon in polar coordinates with the
electron at the origin.\Ref\ac{See Y. Aharonov and A. Casher, {\it Phys. Rev.
Lett.} {\bf 33}, 319 (1984).  The Aharonov-Casher (AC) effect is dual to the AB
effect:  a neutron interacting with a line of charge is equivalent to an
electron interacting with a fluxon.  Eq. \hamhalf\ is an effective
two-dimensional Hamiltonian for both the AB and AC effects, and all our
conclusions hold for both cases.} Consider a limited time reversal operation
$T$ sending $\vec p_i \rightarrow -\vec p_i$ but leaving $\vec A$ unchanged.
$T$ sends $(\vec p_i \pm e\vec A)^2 \rightarrow (\vec p_i \mp e\vec A)^2$;
since $\vec A \ne -\vec A$, $T$ seems not to be a symmetry of $H$. However, for
the special case of a {\it half-fluxon}, the difference between $\vec A$ and
$-\vec A$ amounts to a pure gauge transformation: $\vec A = -\vec A + \nabla_2
\Lambda$ with $\Lambda \equiv (\hbar /e) \varphi_{12}$; so $T$ {\it is} a
symmetry of $H$.  Thus there is a gauge in which we can choose the eigenstates
of $H$ real. Suppose $m_2 >> m_1$.  If we apply the Born-Oppenheimer
approximation to Eq. \hamhalf, the effective Hamiltonian for the fluxon will
contain an induced vector potential due to adiabatic transport of a {\it real}
electron wave function, thus it will preserve the time reversal symmetry.  Let
an initial state for the half-fluxon evolve according to this effective
Hamiltonian and move adiabatically around a loop $C$.  The state accumulates a
phase factor which may include a geometric as well as a dynamical phase.  Time
reversal symmetry implies that the state acquires the {\it same} phase factor
if it moves around $C$ in the opposite sense.  The dynamical phase is the same
in the two cases but the {\it geometric phase} $\Phi (C)$ changes sign.  Since
the overall phase factor remains unchanged, we conclude that the geometric
phase acquired by the electron-fluxon system can be only 0 or $\pi$.

     What, then, becomes of the geometrical phase $\Phi (C)$ as we deform the
path $C$?  Let us assume the electronic wave function to be restricted to a
finite region $\cal S$.  Fig. 1 shows a closed fluxon path $C_1$ which lies
completely outside the region $\cal S$ without encircling it. For this path,
the AB phase is zero. Furthermore, Berry's phase also vanishes.  Now let us
gradually distort the path $C_1$  until it becomes a large loop $C_2$ that
encircles the region $\cal S$ without touching it. For this loop Berry's phase
vanishes, but the AB phase is $\Phi_{AB} (C_2 )=\pi$, since all the charge has
been encircled once.\Ref\proof{Let $\Psi (\vec r_1 , \vec r_2 )$ represent a
localized fluxon wave packet encircling the electron without entering $\cal S$.
 Since the wave packet is localized, the wave function $\Psi^\prime$ defined by
$\Psi =e^{(i/2) \varphi_{12} }\Psi^\prime$ is single-valued.  Note
$\Psi^\prime$ solves Eq. \hamhalf\ with $\vec A =0$.  The remaining factor
$e^{(i/2) \varphi_{12} }$ in $\Psi$ yields the phase $\pi$ for a circuit about
$\cal S$.} We can distort $C_1$ into $C_2$ by many steps which enlarge the loop
by an infinitesimal region. Naively, we would expect the phase $\Phi$ of the
loop to vary smoothly from 0 to $\pi$ but, as noted, $\Phi$ can only be $0$ or
$\pi$.  Thus, we conclude that some infinitesimal region contains a ``singular
point'' $\cal P$ so that $\Phi$ jumps when this infinitesimal region is
annexed.  The electronic wave function yields a vector potential that is always
bounded, and so an infinitesimal region cannot lead to a jump in the AB phase.
Therefore, the jump in $\Phi$ is due to Berry's phase.  The significance of
$\cal P$ is clear:  $\cal P$ is a point such that if a half-fluxon is
introduced there, the electron wave function becomes degenerate.  (Only a
degeneracy can cause such a jump in Berry's phase.) The feature that we exhibit
with this indirect argument, namely that such a point $\cal P$ exists (even if
$V(\vec r_1 )\ne V(r_1 )$), would be hard to see from a direct study of
Schr\"odinger's equation.

     Conversely, suppose we suspect that two states become degenerate at a
point $\cal P$.  Near $\cal P$, we can truncate the Hilbert space for the
system to the subspace spanned by the two states, and write the effective
Hamiltonian as a sum of Pauli matrices (plus a constant) $H_0(x,y) + H_1
(x,y)\sigma_1 + H_2 (x,y)\sigma_2 + H_3(x,y)\sigma_3$. For a generic fluxon,
the degeneracy condition involves {\it three} equations with two parameters
$x,y$ (the coordinates of the fluxon), so that there are no solutions. However,
for the special case of a half-fluxon, the eigenstates and thus the effective
Hamiltonian can always be chosen real. Then $H_2 (x,y)$ vanishes.  The
degeneracy point ${\cal P} =(x^*,y^*)$ is fixed by requiring
$H_1(x^*,y^*)=H_3(x^*,y^*)=0$; these two equations naturally lead to isolated
points of degeneracy.

     The actual location of $\cal P$ depends on the state $\Psi_0$ and relevant
potential. When the potential is spherically symmetric, $V(\vec r) =V(r)$, the
point $\cal P$ corresponds to a fluxon through the $z$-axis. The Hamiltonian
retains azimuthal symmetry.  If states depend on $\varphi$ as $e^{im\varphi}$
for integer $m$, introducing the fluxon is equivalent to shifting the angular
momentum $L_z$ by half a unit:  $-i(\partial /\partial \varphi ) \rightarrow
-i(\partial / \partial \varphi) -1/2$ or $L_z\rightarrow L_z-\hbar /2$.
Initially the energy is proportional to $m^2$. All the energy levels are doubly
degenerate except for the ground state. The shift $m\rightarrow m'=m-1/2$
rearranges $all$ the levels into degenerate pairs. In particular, the ground
state $m=0$ becomes degenerate with $m=1$ (since $m=0\ra m'=-1/2$ and $m=1\ra
m'=1/2$). This degeneracy occurs only for a half-fluxon.

     There could be any odd number of degeneracy points. Indeed, consider the
$m' =\pm 3/2$ (degenerate) states of a rotationally symmetric potential with a
half-fluxon at the center.  By adding a perturbation $V^\prime = \lambda \cos
3\varphi$ which connects these two states, the degeneracy is lifted.  To
restore the degeneracy, we must move the half-fluxon away from the origin.  The
problem is now invariant under rotations of $2 \pi /3$ and so, by symmetry,
there will be three degeneracy points.  A similar argument with $m' =\pm
(2k+1)/2$ and $V^\prime = \lambda \cos (2k+1)\varphi$ leads to $2k+1$
symmetrically situated points.\Ref\k{By contrast, $V^\prime =\lambda \cos (
2k\varphi )$ does not connect any degenerate pair and this argument fails -- as
it must, since the number of degeneracy points cannot be even.}

     Let us now determine the phase collected by an atom which slowly moves in
the presence of {\it two} semifluxons. When the undisturbed ground state is
spherically symmetric and the fluxons are fixed, we can map this problem to an
equivalent one, replacing the spherical charge distribution by a point charge
located at its center, and the fluxons by ``shadow'' fluxons. The shadow
fluxons are defined as points such that when the center of the atom coincides
with one of them, a degeneracy results.  The winding number of the path of the
point charge around the shadow fluxons gives the phase accumulated by the atom.
Consider two straight and parallel semifluxon lines situated a distance $L$
apart. Two extreme cases are easily solved. When the distance between the
fluxons is much larger than the size of the atom, we can move the atom in the
vicinity of one of the fluxons without the electron cloud crossing the other
fluxon. In this case the atom collects a phase of $\pi$ each time its center
encircles the fluxon, exactly as if the other fluxon were not present. The
``shadow'' fluxons coincide therefore with the original fluxons. On the other
hand, for $L=0$ the two semifluxons are at the same point, adding up to an
integer fluxon with no effect on the energy levels of the electron, and
therefore no ``shadow'' fluxons can exist. When the fluxons are slightly
separated, they do affect the energy.  However, by continuity, an infinitesimal
separation of the fluxons cannot produce a degeneracy; rather, a minimal
distance $L^* >0$ is required. Thus we arrive at the conclusion that in an
adiabatic quantum process (say, an atom in a specific state moving slowly) the
geometric phase due to two half-fluxons will always be zero once their
separation $L$ is less than some $L^* >0$.

     We may now interpolate between $L=L^*$ and $L\to\infty$. Instead of
considering the atom as moving by fixed semifluxons, let us fix the atom and
one semifluxon and allow the second semifluxon to move. Let the center of the
atom be at ${\cal O}$ and a fluxon $F_1$ at $\p_1$, and let us determine the
phase accumulated by a second fluxon $F_2$ as it slowly moves along various
closed paths (Fig. 2).  Again, this phase can only be 0 or $\pi$; thus there
must be a point $\p_2$ such that when $F_2$ encircles $\p_2$, the phase jumps
by $\pi$. Insertion of the fluxon $F_2$ at the point $\p_2$ produces a
degeneracy. The connection with the ``shadows'' is that here the point ${\cal
O}$ corresponds to a shadow fluxon.  Let us assume that the points $\p_1$ and
$\p_2$ are related by a continuous function. By symmetry, $\p_1$, $\p_2$ and
${\cal O}$ must form a straight line. We claim that $\p_1$ and $\p_2$ lie on
opposite sides of ${\cal O}$. Let us examine $\p_1$ as a function of $\p_2$. If
$\p_2$ is located in the region where the wave function vanishes, $\p_1$ must
be situated at the atom's center ${\cal O}$. As $\p_2$ enters the electron
cloud and moves towards ${\cal O}$, $\p_1$ must move either towards $\p_2$  or
in the opposite direction.  The first possibility must be discarded:  in this
case either $\p_1$ and $\p_2$ will collide, or $\p_1$ will reverse direction
and eventually return to ${\cal O}$ to avoid collision with $\p_2$. Both
alternatives are inconsistent. If the two half-fluxons collide, they form an
integer fluxon with no degeneracy. If $\p_1$ reverses direction, we obtain an
``accidental" degeneracy with $\p_1$ at ${\cal O}$ and $\p_2$ inside the
electron cloud, where a degeneracy cannot arise.\Ref\nl{P. A. M. Dirac, {\it
Proc. Roy. Soc.} {\bf A133}, 60 (1931) showed that the wave function for a
charged particle, in the presence of a magnetic monopole, must have a line of
zeros extending from the monopole to an antipole or to infinity.  An analogous
argument shows that, in the presence of a semifluxon, a nondegenerate state
$\Psi_0$ must have a line of zeros extending from the semifluxon to another
semifluxon or to infinity.  (A {\it surface} of zeros issues from a
three-dimensional fluxon, but we refer to a {line} for the effective
two-dimensional problem.)  In the ground state only one such null line issues
from each semifluxon. We can choose the vector potential $\vec A$ to be
singular on the null line and zero elsewhere.  Approaching the ``accidental"
degeneracy configuration, the semifluxons carry null lines which, by symmetry,
must lie along their common line. A null line can connect the fluxons, or two
null lines can issue from them in opposite directions. The corresponding wave
functions are not degenerate because the two null lines are more constraining
than one.} As claimed, then, the points $\p_1$ and $\p_2$ lie on opposite sides
of ${\cal O}$.  Thus, the shadow fluxon associated with each semifluxon is
shifted towards the other semifluxon.

     It is amusing to consider various patterns of half-fluxons and resulting
shadows. Even in the case of a single half-fluxon, the shadow need not coincide
with the original, if the fluxon line is not straight. For two half-fluxon
lines intersecting at an acute angle, we expect to find shadow fluxons in the
plane of the half-fluxons, located near the latter but shifted towards a more
acute angle.  Then from continuity, we expect ``hyperbolic" shadow fluxons as
shown in Fig. 3.  For the case of $n$ half-fluxon lines in a plane intersecting
symmetrically at one point, the shadow fluxons will be identical with the
half-fluxons and will induce simple degeneracies. The intersection could be a
point of higher degeneracy.

     Finally, we discuss the case of $N$ semifluxons and an electron cloud of
arbitrary shape. For simplicity we consider a two-dimensional problem. The set
of points $(\p_1,\p_2,\dots,\p_N)$ such that if in each of them a semifluxon is
introduced, the initial wave function of the electron becomes degenerate,
constitutes a $(2N-2)$-dimensional hypersurface $\Sigma$. Indeed, for {\it any}
given points $\p_1,\p_2,\dots,\p_{N-1}$ there always exists at least one
corresponding point $\p_N$. As proof, we introduce a semifluxon in each of the
$N-1$ points $\p_1,\dots,\p_{N-1}$ and consider the phase accumulated by the
$N$-th semifluxon as it takes various paths. Similar arguments to those above
for one and two semifluxons lead to the conclusion that at some point $\p_N$,
the Berry phase jumps by $\pi$. Introducing semifluxons at
$\p_1,\dots,\p_{N-1}, \p_N$ therefore results in a degeneracy.

     We may describe the locations of the $N$ fluxons by a point $(x_1,
y_1,\dots, x_N, y_N)$. To every configuration of $N-1$ fluxons there
corresponds a location $x_N,y_N$ where the $N$-th fluxon induces a degeneracy:
$$
x_N=f(x_1, y_1,...,x_{N-1}, y_{N-1})  \ \ \
y_N=g(x_1, y_1,...,x_{N-1}, y_{N-1}).
\eqn\xy $$
Eq. \xy\ then defines the $(2N-2)$-dimensional hypersurface $\Sigma$. Suppose
that $N$ semifluxons move slowly and after a certain time all return to their
initial positions. What is the topological phase in this case? The fluxons
describe a closed path $C$ in the $2N$-dimensional space  $(x_1, y_1,\dots,x_N,
y_N)$. The phase accumulated by the fluxons as they move through the charge
distribution is simply $n\pi$, where $n$ is the winding number of the closed
path ${\cal C}$ around $\Sigma$.
\bigskip
\centerline{\bf \quad ACKNOWLEDGEMENTS}
\medskip
We thank Aharon Casher for discussions. This research was supported in part by
grant 425-91-1 of the Basic Research Foundation, administered by the Israel
Academy of Sciences and Humanities, by the Program in Alternative Thinking at
Tel-Aviv University, and by grants PHY 8807812 and PHY 90-8936 of the National
Science Foundation.
\vfill\eject

\refout

\vfill\eject

\centerline{\bf \quad FIGURE CAPTIONS}

Fig 1.  The shaded region $S$ indicates where the wave function is non-zero,
and $C_1$ and $C_2$ are limiting paths. Insertion of a half-fluxon at point $P$
induces degeneracy.

Fig. 2.  The center of the atom is ${\cal O}$; at points $P_1$, $P_2$
simultaneous insertion of half-fluxons induces degeneracy. $C_1$ and $C_2$ are
limiting paths of the half-fluxon $F_2$.

Fig. 3.  Two half-fluxons intersecting at an acute angle and the resulting
shadow fluxons.

\vfill\eject

\bye%
ú rohrlich TAUPHY 12/14/93
§Daniel Rohrlich     daniel@taunivm      12/14/93 fluxet.tex
======================================================================== 337
Return-Path: <@TAUNIVM.TAU.AC.IL:rohrlich@TAUPHY.TAU.AC.IL>
Received: from TAUNIVM (NJE origin SMTP@TAUNIVM) by TAUNIVM.TAU.AC.IL (LMail
          V1.1d/1.7f) with BSMTP id 2472; Tue, 14 Dec 1993 23:31:57 +0200
Received: from tauphy by VM.TAU.AC.IL (IBM VM SMTP V2R1) with TCP;
   Tue, 14 Dec 93 23:31:53 IST
Date: Tue, 14 Dec 1993 23:31:54 +0300
{}From: rohrlich@tauphy.tau.ac.il
To: daniel@taunivm
Subject: fluxet.tex
X-VMS-To: SMTP

\line{\hfil TAUP 2110-93}

\def \ra {\rightarrow}
\def \p {{\cal P}}

\medskip
\titlepage{A-B AND BERRY PHASES FOR A QUANTUM CLOUD OF CHARGE}
\vskip 2 true cm
\centerline{
{{Yakir Aharonov,\footnote{a}{School of Physics and Astronomy,
Tel-Aviv University, Ramat-Aviv 69978 Israel}}\footnote{b}{also
Department of Physics and Astronomy, University of South Carolina,
Columbia SC 29208}}
{Sidney Coleman,\footnote{c}{Lyman Laboratory, Harvard University,
Cambridge, MA 02138}}
{Alfred S. Goldhaber,\footnote{d}{Institute for Theoretical Physics,
State University of New York, Stony Brook, NY 11794-3840}}
{Shmuel Nussinov,$^a$}
}
\centerline{
{Sandu Popescu,\footnote{e}
{Service de Physique Theorique, Universit\'e Libre de Bruxelles,
Campus Plaine, C.P. 225, Boulevard du Triomphe, B-1050 Brussels,
Belgium}}
{Benni Reznik,$^a$}
{Daniel Rohrlich,$^a$}
and {Lev Vaidman$^a$}
}
\vskip 4 true cm

\abstract
We investigate the phase accumulated by a charged particle in an extended
quantum state as it encircles one or more magnetic fluxons, each carrying half
a flux unit. A simple, essentially topological analysis reveals an interplay
between the Aharonov-Bohm phase and Berry's phase.
\vfill
\eject

     The Aharonov-Bohm (AB) phase\Ref\ab{Y.~Aharonov and D.~ Bohm, {\it Phys.
Rev.} {\bf 115}, 485 (1959).} $\Phi_{AB}= (q /\hbar )\oint_C\vec A\cdot d\vec
r$ collected by a charge $q$, moving in a closed path $C$ about a line of
magnetic flux $\phi$, is purely topological: $\Phi_{AB}= 2\pi n (q/e)(\phi
/\phi_0)$, with $n$ the winding number of $C$ around the fluxon, $e$ the
elementary charge, and $\phi_0$ the corresponding flux unit $\phi_0={2\pi\hbar
/e}$. The AB phase is independent of the shape of the path $C$ and of the
history of motion along it.  If the charge is not pointlike, or the fluxon is
not linelike, they may overlap; what then happens to the AB phase? As long as
the charge and flux are distributed classically, the answer is straightforward:
 a system of charges moving in a closed path through a classical magnetic field
collects an AB phase $\Phi_{AB}= (1/\hbar ) \sum q_i\phi_{i}$, with $\phi_{i}$
the flux enclosed by the path of the $i$-th charge. Here, however, we consider
charges distributed by quantum smearing. The phase of a {\it quantum} charge is
no simple sum over the undeformed charge distribution.  If we compute it via
the Born-Oppenheimer approximation, we find a remarkable interplay between the
AB phase and Berry's phase that determines the overall topological phase.

     Consider a single electron bound to a heavy ``nucleus'' (assumed neutral,
for simplicity) in the presence of an infinitely long flux line. Both the
nucleus and the fluxon may move. If the fluxon makes a closed path around the
nucleus, it may encircle some parts of the quantum charge distribution (the
electron ``cloud'') and not others.  Suppose that the time in which the fluxon
crosses the ``atom", multiplied by typical electronic frequencies, is much
smaller than 1.  In this limit, the initial electronic wave function $\Psi_0$
transforms into $\Psi_1 = e^{i\Phi_{AB}}\Pi_C \Phi_0 + (1-\Pi_C ) \Phi_0$,
where $\Pi_C$ projects onto the part of $\Psi_0$ that the fluxon encircles.
The electron has no time to move between the two parts of $\Psi_0$. But, except
in this limit, we cannot assign parts of the wave function distinct phases.
Consider now the opposite limit, of adiabatic motion. In this limit, another
phase effect comes into play.  Berry's phase\Ref\be{M.~V.~Berry, {\it Proc.
Roy. Soc. London} {\bf A392}, 45 (1984).} arises when parameters for a quantum
system vary adiabatically in a closed path. Applying the Born-Oppenheimer
approximation to the fluxon-atom system, and for definiteness fixing the
nucleus, we obtain {\it both} an AB phase and a Berry phase. An AB phase arises
from motion of the fluxon with respect to the instantaneous charge
distribution, while a Berry phase arises from rearrangement of the electronic
wave function.  There is a subtle interplay of these two phases, which is
purely topological for special values of the flux $\phi$ carried by the fluxon.
 For example, when $\phi=n\phi_0$, the phases completely cancel \Ref\prep{B.~
Reznik and Y.~Aharonov, {\it Phys. Lett.} {\bf B315}, 386 (1993).} as expected
since the fluxon is a pure gauge artifact. Here, we consider the more
interesting case of fluxons carrying half a flux unit (``half-fluxons" or
``semifluxons"). Topological analysis, with no computations, reveals the
interplay of the AB phase and Berry's phase.

     Let us begin with the electron (mass $m_1$) bound at the origin with a
potential $V(\vec r_1 )$ and the fluxon (mass $m_2$) free to move but
constrained to remain parallel to the $z$-axis.  The Hamiltonian is
$$
H= {(\vec p_1 +e\vec A)^2\over 2m_1 }+V(\vec r_1 )
+ {(\vec p_2 -e\vec A)^2\over 2m_2 }~~~~;
\eqn\hamhalf $$
for a half-fluxon we take $\vec A = (\hbar /2e) \vec \nabla_2 \varphi_{12}$
with $\varphi_{12}$ the angle of the fluxon in polar coordinates with the
electron at the origin.\Ref\ac{See Y. Aharonov and A. Casher, {\it Phys. Rev.
Lett.} {\bf 33}, 319 (1984).  The Aharonov-Casher (AC) effect is dual to the AB
effect:  a neutron interacting with a line of charge is equivalent to an
electron interacting with a fluxon.  Eq. \hamhalf\ is an effective
two-dimensional Hamiltonian for both the AB and AC effects, and all our
conclusions hold for both cases.} Consider a limited time reversal operation
$T$ sending $\vec p_i \rightarrow -\vec p_i$ but leaving $\vec A$ unchanged.
$T$ sends $(\vec p_i \pm e\vec A)^2 \rightarrow (\vec p_i \mp e\vec A)^2$;
since $\vec A \ne -\vec A$, $T$ seems not to be a symmetry of $H$. However, for
the special case of a {\it half-fluxon}, the difference between $\vec A$ and
$-\vec A$ amounts to a pure gauge transformation: $\vec A = -\vec A + \nabla_2
\Lambda$ with $\Lambda \equiv (\hbar /e) \varphi_{12}$; so $T$ {\it is} a
symmetry of $H$.  Thus there is a gauge in which we can choose the eigenstates
of $H$ real. Suppose $m_2 >> m_1$.  If we apply the Born-Oppenheimer
approximation to Eq. \hamhalf, the effective Hamiltonian for the fluxon will
contain an induced vector potential due to adiabatic transport of a {\it real}
electron wave function, thus it will preserve the time reversal symmetry.  Let
an initial state for the half-fluxon evolve according to this effective
Hamiltonian and move adiabatically around a loop $C$.  The state accumulates a
phase factor which may include a geometric as well as a dynamical phase.  Time
reversal symmetry implies that the state acquires the {\it same} phase factor
if it moves around $C$ in the opposite sense.  The dynamical phase is the same
in the two cases but the {\it geometric phase} $\Phi (C)$ changes sign.  Since
the overall phase factor remains unchanged, we conclude that the geometric
phase acquired by the electron-fluxon system can be only 0 or $\pi$.

     What, then, becomes of the geometrical phase $\Phi (C)$ as we deform the
path $C$?  Let us assume the electronic wave function to be restricted to a
finite region $\cal S$.  Fig. 1 shows a closed fluxon path $C_1$ which lies
completely outside the region $\cal S$ without encircling it. For this path,
the AB phase is zero. Furthermore, Berry's phase also vanishes.  Now let us
gradually distort the path $C_1$  until it becomes a large loop $C_2$ that
encircles the region $\cal S$ without touching it. For this loop Berry's phase
vanishes, but the AB phase is $\Phi_{AB} (C_2 )=\pi$, since all the charge has
been encircled once.\Ref\proof{Let $\Psi (\vec r_1 , \vec r_2 )$ represent a
localized fluxon wave packet encircling the electron without entering $\cal S$.
 Since the wave packet is localized, the wave function $\Psi^\prime$ defined by
$\Psi =e^{(i/2) \varphi_{12} }\Psi^\prime$ is single-valued.  Note
$\Psi^\prime$ solves Eq. \hamhalf\ with $\vec A =0$.  The remaining factor
$e^{(i/2) \varphi_{12} }$ in $\Psi$ yields the phase $\pi$ for a circuit about
$\cal S$.} We can distort $C_1$ into $C_2$ by many steps which enlarge the loop
by an infinitesimal region. Naively, we would expect the phase $\Phi$ of the
loop to vary smoothly from 0 to $\pi$ but, as noted, $\Phi$ can only be $0$ or
$\pi$.  Thus, we conclude that some infinitesimal region contains a ``singular
point'' $\cal P$ so that $\Phi$ jumps when this infinitesimal region is
annexed.  The electronic wave function yields a vector potential that is always
bounded, and so an infinitesimal region cannot lead to a jump in the AB phase.
Therefore, the jump in $\Phi$ is due to Berry's phase.  The significance of
$\cal P$ is clear:  $\cal P$ is a point such that if a half-fluxon is
introduced there, the electron wave function becomes degenerate.  (Only a
degeneracy can cause such a jump in Berry's phase.) The feature that we exhibit
with this indirect argument, namely that such a point $\cal P$ exists (even if
$V(\vec r_1 )\ne V(r_1 )$), would be hard to see from a direct study of
Schr\"odinger's equation.

     Conversely, suppose we suspect that two states become degenerate at a
point $\cal P$.  Near $\cal P$, we can truncate the Hilbert space for the
system to the subspace spanned by the two states, and write the effective
Hamiltonian as a sum of Pauli matrices (plus a constant) $H_0(x,y) + H_1
(x,y)\sigma_1 + H_2 (x,y)\sigma_2 + H_3(x,y)\sigma_3$. For a generic fluxon,
the degeneracy condition involves {\it three} equations with two parameters
$x,y$ (the coordinates of the fluxon), so that there are no solutions. However,
for the special case of a half-fluxon, the eigenstates and thus the effective
Hamiltonian can always be chosen real. Then $H_2 (x,y)$ vanishes.  The
degeneracy point ${\cal P} =(x^*,y^*)$ is fixed by requiring
$H_1(x^*,y^*)=H_3(x^*,y^*)=0$; these two equations naturally lead to isolated
points of degeneracy.

     The actual location of $\cal P$ depends on the state $\Psi_0$ and relevant
potential. When the potential is spherically symmetric, $V(\vec r) =V(r)$, the
point $\cal P$ corresponds to a fluxon through the $z$-axis. The Hamiltonian
retains azimuthal symmetry.  If states depend on $\varphi$ as $e^{im\varphi}$
for integer $m$, introducing the fluxon is equivalent to shifting the angular
momentum $L_z$ by half a unit:  $-i(\partial /\partial \varphi ) \rightarrow
-i(\partial / \partial \varphi) -1/2$ or $L_z\rightarrow L_z-\hbar /2$.
Initially the energy is proportional to $m^2$. All the energy levels are doubly
degenerate except for the ground state. The shift $m\rightarrow m'=m-1/2$
rearranges $all$ the levels into degenerate pairs. In particular, the ground
state $m=0$ becomes degenerate with $m=1$ (since $m=0\ra m'=-1/2$ and $m=1\ra
m'=1/2$). This degeneracy occurs only for a half-fluxon.

     There could be any odd number of degeneracy points. Indeed, consider the
$m' =\pm 3/2$ (degenerate) states of a rotationally symmetric potential with a
half-fluxon at the center.  By adding a perturbation $V^\prime = \lambda \cos
3\varphi$ which connects these two states, the degeneracy is lifted.  To
restore the degeneracy, we must move the half-fluxon away from the origin.  The
problem is now invariant under rotations of $2 \pi /3$ and so, by symmetry,
there will be three degeneracy points.  A similar argument with $m' =\pm
(2k+1)/2$ and $V^\prime = \lambda \cos (2k+1)\varphi$ leads to $2k+1$
symmetrically situated points.\Ref\k{By contrast, $V^\prime =\lambda \cos (
2k\varphi )$ does not connect any degenerate pair and this argument fails -- as
it must, since the number of degeneracy points cannot be even.}

     Let us now determine the phase collected by an atom which slowly moves in
the presence of {\it two} semifluxons. When the undisturbed ground state is
spherically symmetric and the fluxons are fixed, we can map this problem to an
equivalent one, replacing the spherical charge distribution by a point charge
located at its center, and the fluxons by ``shadow'' fluxons. The shadow
fluxons are defined as points such that when the center of the atom coincides
with one of them, a degeneracy results.  The winding number of the path of the
point charge around the shadow fluxons gives the phase accumulated by the atom.
Consider two straight and parallel semifluxon lines situated a distance $L$
apart. Two extreme cases are easily solved. When the distance between the
fluxons is much larger than the size of the atom, we can move the atom in the
vicinity of one of the fluxons without the electron cloud crossing the other
fluxon. In this case the atom collects a phase of $\pi$ each time its center
encircles the fluxon, exactly as if the other fluxon were not present. The
``shadow'' fluxons coincide therefore with the original fluxons. On the other
hand, for $L=0$ the two semifluxons are at the same point, adding up to an
integer fluxon with no effect on the energy levels of the electron, and
therefore no ``shadow'' fluxons can exist. When the fluxons are slightly
separated, they do affect the energy.  However, by continuity, an infinitesimal
separation of the fluxons cannot produce a degeneracy; rather, a minimal
distance $L^* >0$ is required. Thus we arrive at the conclusion that in an
adiabatic quantum process (say, an atom in a specific state moving slowly) the
geometric phase due to two half-fluxons will always be zero once their
separation $L$ is less than some $L^* >0$.

     We may now interpolate between $L=L^*$ and $L\to\infty$. Instead of
considering the atom as moving by fixed semifluxons, let us fix the atom and
one semifluxon and allow the second semifluxon to move. Let the center of the
atom be at ${\cal O}$ and a fluxon $F_1$ at $\p_1$, and let us determine the
phase accumulated by a second fluxon $F_2$ as it slowly moves along various
closed paths (Fig. 2).  Again, this phase can only be 0 or $\pi$; thus there
must be a point $\p_2$ such that when $F_2$ encircles $\p_2$, the phase jumps
by $\pi$. Insertion of the fluxon $F_2$ at the point $\p_2$ produces a
degeneracy. The connection with the ``shadows'' is that here the point ${\cal
O}$ corresponds to a shadow fluxon.  Let us assume that the points $\p_1$ and
$\p_2$ are related by a continuous function. By symmetry, $\p_1$, $\p_2$ and
${\cal O}$ must form a straight line. We claim that $\p_1$ and $\p_2$ lie on
opposite sides of ${\cal O}$. Let us examine $\p_1$ as a function of $\p_2$. If
$\p_2$ is located in the region where the wave function vanishes, $\p_1$ must
be situated at the atom's center ${\cal O}$. As $\p_2$ enters the electron
cloud and moves towards ${\cal O}$, $\p_1$ must move either towards $\p_2$  or
in the opposite direction.  The first possibility must be discarded:  in this
case either $\p_1$ and $\p_2$ will collide, or $\p_1$ will reverse direction
and eventually return to ${\cal O}$ to avoid collision with $\p_2$. Both
alternatives are inconsistent. If the two half-fluxons collide, they form an
integer fluxon with no degeneracy. If $\p_1$ reverses direction, we obtain an
``accidental" degeneracy with $\p_1$ at ${\cal O}$ and $\p_2$ inside the
electron cloud, where a degeneracy cannot arise.\Ref\nl{P. A. M. Dirac, {\it
Proc. Roy. Soc.} {\bf A133}, 60 (1931) showed that the wave function for a
charged particle, in the presence of a magnetic monopole, must have a line of
zeros extending from the monopole to an antipole or to infinity.  An analogous
argument shows that, in the presence of a semifluxon, a nondegenerate state
$\Psi_0$ must have a line of zeros extending from the semifluxon to another
semifluxon or to infinity.  (A {\it surface} of zeros issues from a
three-dimensional fluxon, but we refer to a {line} for the effective
two-dimensional problem.)  In the ground state only one such null line issues
from each semifluxon. We can choose the vector potential $\vec A$ to be
singular on the null line and zero elsewhere.  Approaching the ``accidental"
degeneracy configuration, the semifluxons carry null lines which, by symmetry,
must lie along their common line. A null line can connect the fluxons, or two
null lines can issue from them in opposite directions. The corresponding wave
functions are not degenerate because the two null lines are more constraining
than one.} As claimed, then, the points $\p_1$ and $\p_2$ lie on opposite sides
of ${\cal O}$.  Thus, the shadow fluxon associated with each semifluxon is
shifted towards the other semifluxon.

     It is amusing to consider various patterns of half-fluxons and resulting
shadows. Even in the case of a single half-fluxon, the shadow need not coincide
with the original, if the fluxon line is not straight. For two half-fluxon
lines intersecting at an acute angle, we expect to find shadow fluxons in the
plane of the half-fluxons, located near the latter but shifted towards a more
acute angle.  Then from continuity, we expect ``hyperbolic" shadow fluxons as
shown in Fig. 3.  For the case of $n$ half-fluxon lines in a plane intersecting
symmetrically at one point, the shadow fluxons will be identical with the
half-fluxons and will induce simple degeneracies. The intersection could be a
point of higher degeneracy.

     Finally, we discuss the case of $N$ semifluxons and an electron cloud of
arbitrary shape. For simplicity we consider a two-dimensional problem. The set
of points $(\p_1,\p_2,\dots,\p_N)$ such that if in each of them a semifluxon is
introduced, the initial wave function of the electron becomes degenerate,
constitutes a $(2N-2)$-dimensional hypersurface $\Sigma$. Indeed, for {\it any}
given points $\p_1,\p_2,\dots,\p_{N-1}$ there always exists at least one
corresponding point $\p_N$. As proof, we introduce a semifluxon in each of the
$N-1$ points $\p_1,\dots,\p_{N-1}$ and consider the phase accumulated by the
$N$-th semifluxon as it takes various paths. Similar arguments to those above
for one and two semifluxons lead to the conclusion that at some point $\p_N$,
the Berry phase jumps by $\pi$. Introducing semifluxons at
$\p_1,\dots,\p_{N-1}, \p_N$ therefore results in a degeneracy.

     We may describe the locations of the $N$ fluxons by a point $(x_1,
y_1,\dots, x_N, y_N)$. To every configuration of $N-1$ fluxons there
corresponds a location $x_N,y_N$ where the $N$-th fluxon induces a degeneracy:
$$
x_N=f(x_1, y_1,...,x_{N-1}, y_{N-1})  \ \ \
y_N=g(x_1, y_1,...,x_{N-1}, y_{N-1}).
\eqn\xy $$
Eq. \xy\ then defines the $(2N-2)$-dimensional hypersurface $\Sigma$. Suppose
that $N$ semifluxons move slowly and after a certain time all return to their
initial positions. What is the topological phase in this case? The fluxons
describe a closed path $C$ in the $2N$-dimensional space  $(x_1, y_1,\dots,x_N,
y_N)$. The phase accumulated by the fluxons as they move through the charge
distribution is simply $n\pi$, where $n$ is the winding number of the closed
path ${\cal C}$ around $\Sigma$.
\bigskip
\centerline{\bf \quad ACKNOWLEDGEMENTS}
\medskip
We thank Aharon Casher for discussions. This research was supported in part by
grant 425-91-1 of the Basic Research Foundation, administered by the Israel
Academy of Sciences and Humanities, by the Program in Alternative Thinking at
Tel-Aviv University, and by grants PHY 8807812 and PHY 90-8936 of the National
Science Foundation.
\vfill\eject

\refout

\vfill\eject

\centerline{\bf \quad FIGURE CAPTIONS}

Fig 1.  The shaded region $S$ indicates where the wave function is non-zero,
and $C_1$ and $C_2$ are limiting paths. Insertion of a half-fluxon at point $P$
induces degeneracy.

Fig. 2.  The center of the atom is ${\cal O}$; at points $P_1$, $P_2$
simultaneous insertion of half-fluxons induces degeneracy. $C_1$ and $C_2$ are
limiting paths of the half-fluxon $F_2$.

Fig. 3.  Two half-fluxons intersecting at an acute angle and the resulting
shadow fluxons.

\vfill\eject

\bye